\documentstyle[aps]{revtex}
\newcommand{\be}{\begin{equation}}
\newcommand{\ee}{\end{equation}}
\newcommand{\bea}{\begin{eqnarray}}
\newcommand{\eea}{\end{eqnarray}}

\begin{document}
\bibliographystyle{unsrt}

\baselineskip=13pt
\title{Double Phase Slips and Bound Defect Pairs in Parametrically Driven Waves\footnote{To 
appear in the Proceedings of the {\it 15th Symposium on Energy Engineering Sciences},
Office of Basic Energy Sciences, Department of Energy, 1997.}
}
\author{Hermann Riecke and Glen D. Granzow}
\address{Department of Engineering Sciences and Applied Mathematics\\
Northwestern University, Evanston, IL 60208, USA}
\maketitle

\begin{abstract}
 Spatio-temporal chaos in parametrically driven waves is investigated in one and
two dimensions using numerical simulations of Ginzburg-Landau equations. 
A regime is identified in which in one dimension the dynamics
are due to double phase slips. In very small systems they are found to arise
through a Hopf bifurcation off a mixed mode. In large systems they can lead to
a state of localized spatio-temporal chaos, which can be understood within the framework
of phase dynamics. In two dimensions the double phase slips are replaced by bound
defect pairs. Our simulations indicate the possibility of an unbinding transition of these
pairs, which is associated with a transition from ordered to disordered defect chaos.
\end{abstract}


\section{Introduction} 
\label{sec:intro}

While low-dimensional chaotic dynamics are quite well understood
this is not the case for chaotic dynamics of high dimension as, for instance,
 spatio-temporal chaos. Spatio-temporal chaos arises in systems in which spatial degrees
of freedom play an important role and the structures are not only chaotic in time
but also in space. These dynamics arise often in pattern-forming systems when
all ordered patterns become unstable. Of particular
interest are systems in which
the chaos is extensive, i.e. systems in which quantities like
the attractor dimension and the number of positive
Lyapunov exponents grow linearly with the system size, so that one can think of them
as a large number of coupled, chaotic entities. 

Spatio-temporal chaos is found in many experimental systems. It has been
extensively studied in convection where a number of different types have been observed.
In the presence of rotation, a classic result is the occurrence of domain
chaos. It is due to the K\"uppers-Lortz instability which renders steady convection rolls unstable
to rolls with a different orientation \cite{KuLo69}. Since the new rolls
 are susceptible to the same instability
a persistent switching of patches of rolls of different orientations is observed 
\cite{HuEc95}.
Very recently spatio-temporal chaos has also been found (without rotation) 
in regimes in which
the convection rolls are in fact linearly stable. Sufficiently large perturbations, however, 
lead to a state of spiral-defect chaos in which spirals and other types of defects dominate
 \cite{MoBo96}. 

Another type of spatio-temporal chaos arises in electroconvection in 
nematic liquid crystals in a traveling wave regime \cite{DeAh96}. Due to the axial
 anisotropy of this system the waves travel only in certain directions relative to the axis
of anisotropy. In the regime in question they travel obliquely to that direction and
because of the reflection symmetries of the system the dynamics is governed by
the competition of waves traveling in 4 directions. The chaotic dynamics arise 
immediately at the onset of convection and is characterized by defects in the various
wave components. Associated with each defect is a suppression of 
the corresponding wave amplitude leading to domains in which one or two of the
wave components dominate. 
 Some understanding of these dynamics and their possible origin has
been obtained within a set of coupled Ginzburg-Landau equations \cite{RiKr97}.
Spatio-temporal chaos  of traveling waves in an isotropic system arise in
 convection in binary mixtures \cite{PoSu96}.
 
A very rich class of pattern-forming systems are parametrically driven waves, the 
standard realization being surface waves on a fluid that are excited by a vertical shaking
of the container at twice the frequency of the waves. Depending on the fluid parameters
and the driving frequency spatially periodic and quasi-periodic patterns of various kinds
have been found as well as transitions to spatio-temporal chaos \cite{KuGo96a}. Strikingly,
very similar phenomena can be found even if the fluid is replaced by a granular medium 
like sand \cite{MeUm95}. 

In the present communication we present theoretical results for the dynamics of parametrically
driven waves in one and two dimensions within the framework of Ginzburg-Landau equations. 
In the one-dimensional analysis we address the important question how to characterize
the behavior of a spatially and temporally chaotic state on length scales that are
much larger than the typical wavelength. The response of 
regular periodic patterns to long-wave perturbations is well understood; in some sense it 
is a dissipative analogue to sound waves in crystals. For steady
patterns the response is typically diffusive whereas for oscillatory patterns it is
propagative. Both can be described by phase equations (or equations for the local wavenumber). 
Here we describe a chaotic state for which the same type of description is possible, i.e.
although the dynamics on short scales is chaotic in space and time, the large-scale behavior
of that state is diffusive. This striking behavior is due to the fact that the 
chaotic state is
driven by {\it double phase slips} as described below. We show that the homogeneously chaotic
state can become diffusively unstable on large scales and separates into arrays of
chaotic and non-chaotic domains very similar to phase separation in mixtures. This provides
a mechanism that can lead to the {\it localization} of the chaotic dynamics in space.

Experimentally, localized spatio-temporal chaos has been
observed in Taylor vortex flow \cite{BaAn86},
Rayleigh-B\'{e}nard convection \cite{CiRu87}, and parametrically excited surface
waves \cite{KuGo96}. So far, the localization mechanism in 
these systems is, however, only poorly understood.

In the second part we present ongoing work on the dynamics that arise in two dimensions
in the regime in which double-phase slips occur in one dimension. As discussed below
it leads to `fluctuating bound defect pairs' . We present evidence that indicates a
transition from an ordered state of defect chaos to a disordered one. This transition appears 
to be associated with an `unbinding' of the defect pairs. 

\section{Ginzburg-Landau Equations for Parametrically Driven Waves}

To obtain a tractable model for parametrically driven waves we consider systems
that exhibit a supercritical Hopf bifurcation to traveling waves, i.e. 
when the control parameter is
increased beyond a certain threshold value
the basic state becomes unstable to small-amplitude traveling waves. This is, for instance,
the case in electroconvection of nematic liquid crystals \cite{DeAh96}\footnote{Convection
in binary mixtures also exhibits a Hopf bifurcation to traveling waves. It is, however, subcritical \cite{PoSu96}
and the waves appear right away with finite amplitude. An analysis of the parametric
forcing of that system is therefore more complicated \cite{RiCr91}.}.
Just below the Hopf bifurcation  the traveling-wave modes are only weakly damped. 
Therefore a small driving is sufficient to
excite standing  waves of small amplitude. Consequently, a weakly nonlinear description is possible
by expanding about the unforced basic state and treating the forcing and the damping
as small perturbations. This leads to two coupled equations for the complex amplitudes
of the traveling-wave components, i.e. physical quantities 
 like the vertical fluid velocity $u$ in the midplane of a convection system are given by 
\begin{equation}
u({\bf r},t)=\epsilon A (X,Y,T)e^{i({\bf q}_c \cdot {\bf r}-\frac{\omega_e}{2}t)} 
      +\epsilon B (X,Y,T)e^{i({\bf q}_c \cdot {\bf r}+\frac{\omega_e}{2}t)}
      +c.c.+o(\epsilon).                                 \label{oneDu}
\end{equation}
The complex amplitudes $A$ and $B$ vary on the slow time and space scales, 
$T=\epsilon^2t$, $X=\epsilon x$, and $Y=\epsilon y$ with ${\bf r}=(x,y)$, 
and $\epsilon \ll 1$.

Using standard symmetry arguments one obtains for the amplitudes $A$ and $B$ 
in one dimension the Ginzburg-Landau equations \cite{RiCr88}
\begin{eqnarray}
\partial_TA+s\partial_XA &=& d\partial_X^2A+aA+bB
 +cA(|A|^2+|B|^2)+gA|B|^2,                               \label{oneDA} \\
\partial_TB-s\partial_XB &=& d^*\partial_X^2B+a^*B+bA
 +c^*B(|A|^2+|B|^2)+g^*B|A|^2.                           \label{oneDB}
\end{eqnarray}
The coefficients in (\ref{oneDA},\ref{oneDB}) are 
complex except for $s$ and $b$, which are real.
The real part $a_r$ of the coefficient of the linear term $a$ gives the linear 
damping of the traveling waves in the absence of the periodic forcing and is
proportional to the distance from the Hopf bifurcation. 
The coefficient of the linear coupling term $b$ gives the amplitude of the 
periodic forcing as can  be seen from the fact that it 
 breaks the continuous time-translation symmetry $t \rightarrow t+\Delta t$ that
implies the transformation $A \rightarrow Ae^{i\Delta t \omega_e/2}$
$B \rightarrow Be^{-i\Delta t \omega_e/2}$.
The imaginary part $a_i$ of the coefficient of the linear term $a$ gives the 
difference between the frequency of the unforced waves and half the forcing 
frequency $\omega_e$.

The same Ginzburg-Landau equations are obtained also for systems that do not exhibit a
Hopf bifurcation, if they have weakly damped traveling wave modes. This is the case for
surface waves on fluids with small viscosity. Then again only a small driving is necessary
to excite the surface waves and one can perform the same kind of expansions. In that
case $a_r <0$ represents the damping of the waves and
 the group velocity parameter $s$ is in general complex, indicating that
the dissipation depends already linearly on the wavenumber. However, all dissipative
terms are small. 
 
In addition to the trivial solution $A=B=0$ (\ref{oneDA},\ref{oneDB}) possess
three types of simple solutions: $|A|=|B|=const$, $|A| \ne |B|$ (both constant), and
$|A|=|B|$ (both time-periodic). We are in particular interested in the first type,
which corresponds in the physical system to standing waves which are phase-locked to the
parametric forcing, i.e. they are excited by the forcing. With increasing $a_r$ they
become unstable to solutions of the second type, which correspond to traveling waves 
as they exist also in the absence of the periodic forcing. Solutions of the third
type correspond to standing waves that are not phase-locked to the forcing. For $g_r<0$
they are unstable to the traveling waves.  

\begin{figure}[htb] 
\begin{picture}(420,160)(0,0)
\put(-75,-60) {\includegraphics{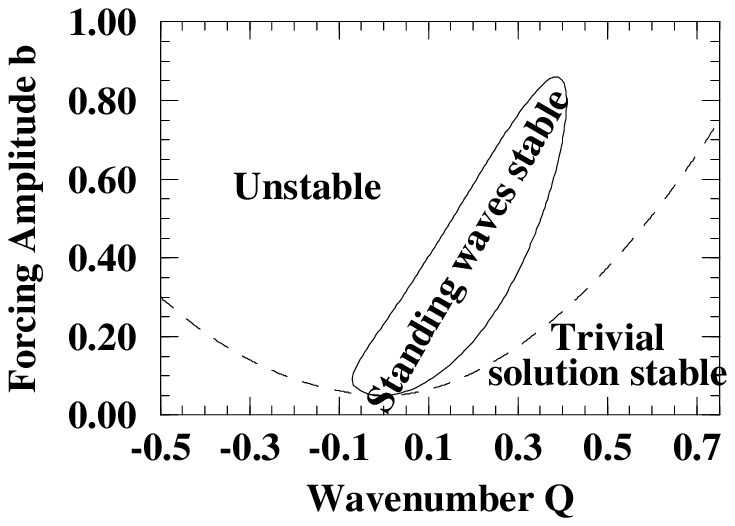}}
\put(165,-60) {\includegraphics{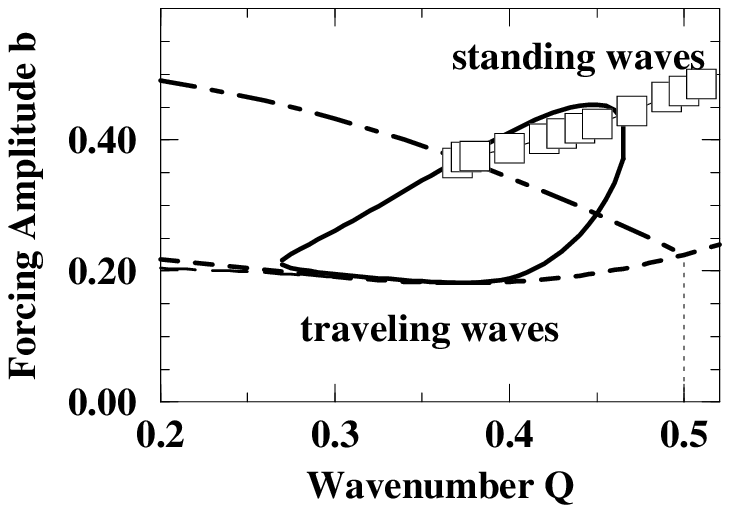}}
\end{picture}
\caption{Linear stability diagram for  
$c=-1+4i$, $d=1+0.5i$, $s=0.2$, $g=-1-12i$. a) $a=-0.05$, b) $a_r=0.25$.
\protect{\label{f:stabi12}}
}
\end{figure}

The response of the phase-locked standing waves to long-wave perturbations
can be described using a phase equation\cite{Ri90a},
\be
\partial_T \phi = D(q) \partial_X^2 \phi \qquad \mbox{ with } q=\partial_X \phi,\label{e:pe}
\ee
which due to the spatial reflection symmetry of the waves is a 
(nonlinear) diffusion equation. 
The diffusion coefficient is not necessarily positive and its sign-change indicates an 
instability of the waves, the Eckhaus instability.
For the one-dimensional case 
the diffusion coefficient was given in \cite{Ri90a}. The resulting stability limits as they are relevant
for the first part of the paper are shown in fig.\ref{f:stabi12}a. The neutral curve, above which the
basic state is unstable to standing-wave perturbations, is given by the dashed line. The solid
line gives the Eckhaus instability of the phase-locked standing waves. In the second
part of the paper we will consider a case in which traveling waves appear in the absence of
forcing, i.e. $a_r>0$. The corresponding stability regions are shown in fig.\ref{f:stabi12}b.
  The neutral curve  is 
given by the dashed line, neutral curve for the appearance of traveling waves 
by a dotted line.  Their Eckhaus instability is denoted by a solid line.
In addition to the Eckhaus instability also a parity-breaking instability arises in which
the standing waves become unstable to traveling waves. It is indicated 
by a dashed-dotted line. 
 Over some range
of parameters the parity-breaking instability is preempted by a mode that arises first
at finite modulation wavenumber (open squares). It emerges from the parity-breaking
instability. The standing waves are 
stable only inside the region marked by the solid lines and the squares.

\section{Dynamics in a Small System}
\label{s:small}

\begin{figure}[htb] 
\begin{picture}(420,200)(0,0)
\put(-75,-60) {\includegraphics{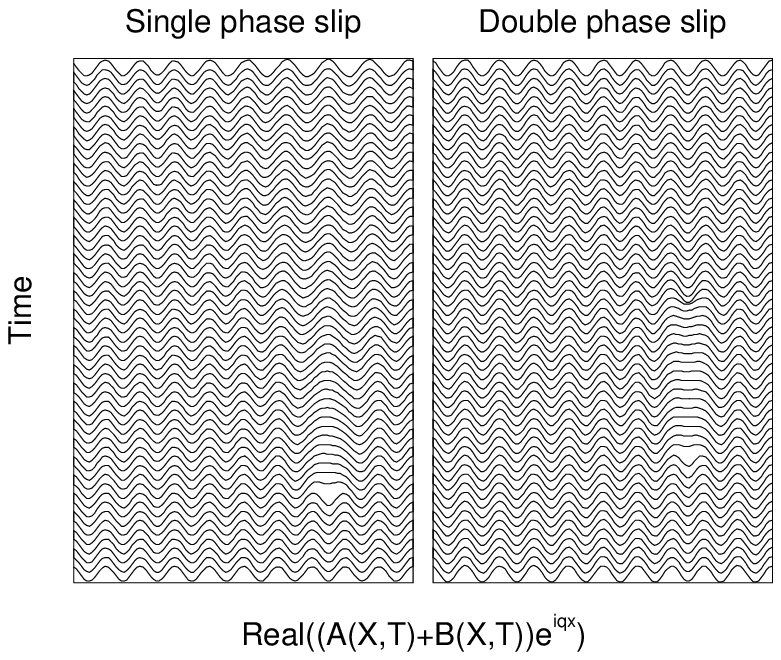}}
\put(175,-50) {\includegraphics{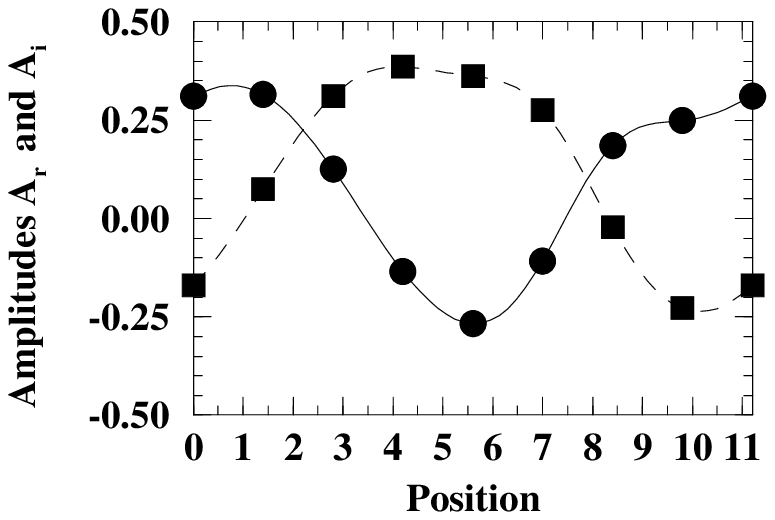}}
\end{picture}
\caption{Space-time diagrams showing a) single phase slip 
for small values of the forcing amplitude (e.g. $b=0.1$), and b) double
phase slip observed for larger values of the forcing amplitude (e.g. $b=0.6$).
Other parameters as in fig.\protect{\ref{f:stabi12}}a.
\protect{\label{f:sps-dps}}
}
\end{figure}

The central new feature of the standing waves to be discussed in this paper is the appearance
of `double phase slips' \cite{GrRi96,GrRi97}. 
Usually the Eckhaus instability leads to a single phase slip which
changes the total phase in the system and through which the wavenumber of the pattern 
jumps from outside the stable band to inside the band. Such an event is shown in fig.\ref{f:sps-dps}a.
While this occurs also in this system when the Eckhaus instability is crossed for weak 
forcing, it is not the case for larger forcing: for $b \approx 0.4$ and above the same perturbation
leads to a double phase slip, which consists of two consecutive phase slips
that undo each other as shown in fig.\ref{f:sps-dps}b. After the double phase slip 
the pattern has the same wave number as before and can undergo the same instability
again leading to persistent dynamics. It can be periodic or irregular as discussed in 
sec.\ref{s:large}. 

\vspace*{-1cm}
\begin{figure}[htb] 
\begin{picture}(420,0)(0,0)
\end{picture}
\caption{Mixed-mode solution for $L=11.2$ \protect{\hfill}
\protect{\label{f:mixedmode}}
}
\end{figure}

\begin{figure}[htb] 
\begin{picture}(420,160)(0,0)
\put(-75,-60) {\includegraphics{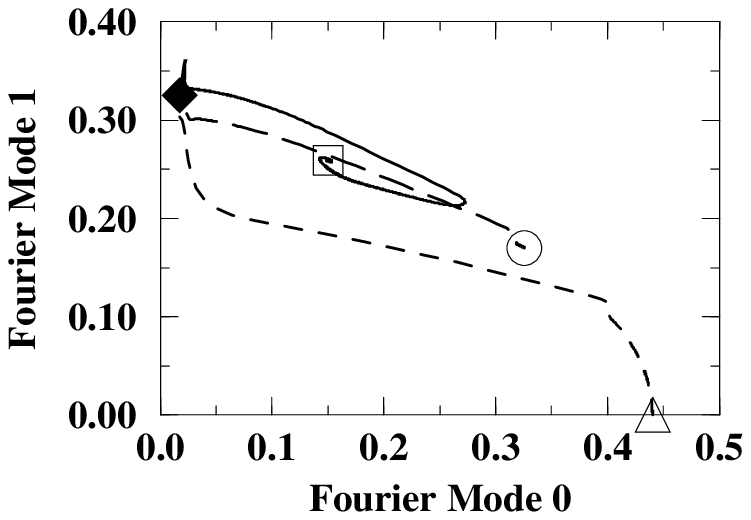}}
\put(165,-60) {\includegraphics{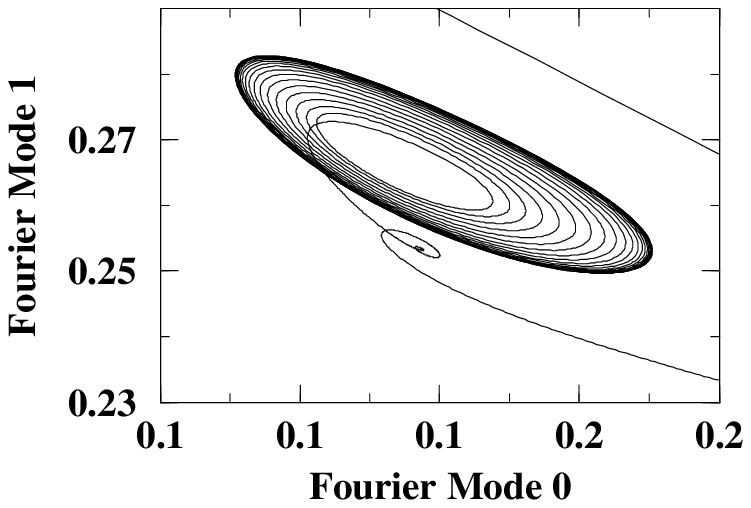}}
\end{picture}
\caption{Phase space projected onto the magnitude of the Fourier modes 0 and 1 for increasing
values of the system length $L$. a) $L=8$ (short dashes), $L=10$ (long dashes), and $L=11.2$
(solid line). Initial condition is marked by a solid diamond, 
the final fixed point by open symbols. b) $L=11.3$.
\protect{\label{f:phasedia}}
}
\end{figure}

To study the origin of the double phase slip we consider a minimal system in which
only the Fourier modes 0 and 1 are important and simulate 
 (\ref{oneDA},\ref{oneDB}) numerically with a pseudospectral code keeping only the 
Fourier modes -4 to +4. Changing the length of the system at fixed $b=1.0$ we find the 
phase portraits shown in fig.\ref{f:phasedia}a and fig.\ref{f:phasedia}b. In each run 
the same pattern very close to the solution with one wavelength
in the system is chosen as initial condition (solid diamond). 
For small length (large wavenumber)  this initial condition
rapidly evolves to the homogeneous solution (short dashes and open triangle). 

When the length is increased  the homogeneous solution becomes unstable and a new stable
fixed point corresponding to a mixed mode involving 
Fourier modes 0 as well as 1 arises (open circle). Consequently the solution converges to that
mixed mode. Fig.\ref{f:mixedmode} shows the real and imaginary parts of the amplitude $A$
of the mixed mode for $L=11.2$.
With increasing $L$ the mixed-mode fixed point moves to the left and the trajectory turns around. For $L=11.2$  the 
mixed-mode fixed point, which was a stable node for smaller $L$, becomes
 a stable spiral point (open square). When $L$ is increased to
 $L=11.3$ the spiral point becomes unstable and generates a stable limit cycle as is
shown in fig.\ref{f:phasedia}b. Closer inspection of that phase portrait
shows that the trajectory follows a three-dimensional path: while it starts
{\it outside} the limit cycle it intersects itself a number of times and eventually approaches
the limit cycle from {\it inside}. This indicates that a model that is to 
capture the mixed mode,
its limit cycle, and the two fixed points corresponding 
to the homogeneous solution and that
with one wavelength has to be at least three-dimensional. For yet larger $L$,
 a second mixed-mode fixed
point appears, which is associated with the Eckhaus instability of the solution with one wavelength. 
It becomes important when the limit cycle grows and eventually 
becomes homoclinic to this additional fixed point around $L=12$.

\section{Dynamics in a Large System}
\label{s:large}
\begin{figure}[htb] 
\begin{picture}(420,160)(0,0)
\put(-75,-60) {\includegraphics{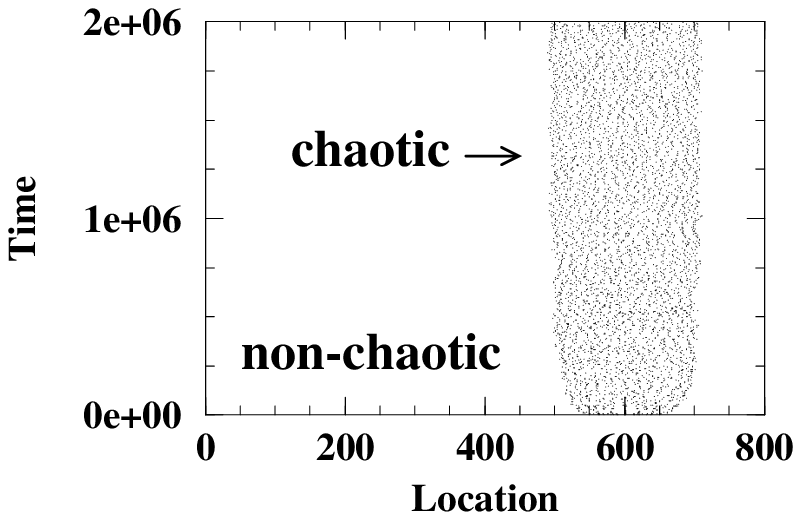}}
\put(165,-60) {\includegraphics{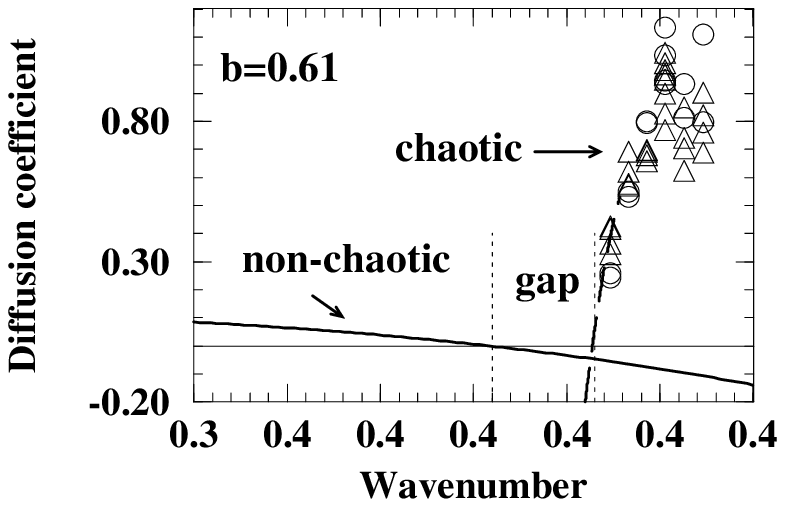}}
\end{picture}
\caption{Localized spatio-temporal chaos for $b=0.6$ (other parameters as in fig.\protect{\ref{f:stabi12}}a);  
the averaged effective wavenumber is $<q>=0.377$). 
\protect{\label{f:locstc}}
}
\end{figure}

While in the small system discussed in sec.\ref{s:small}  only simple dynamics was found,
complex dynamics arises if the double phase slips can occur at more than one location.
 Here we describe results for large systems where a surprisingly simple description of the
large-scale behavior is possible. Our search for such a description was motivated by
states of {\it localized} spatio-temporal chaos \cite{GrRi96,GrRi97}. 
An example is shown in fig.\ref{f:locstc}
where each of the double phase slips is marked as a dot.
An initial perturbation triggers a double phase slip. In its vicinity 
more double phase slips arise and the chaotic
activity starts to spread. However, by $t=50,000$ the width of the chaotic domain stops 
growing and a stable state is reached, in which the chaotic activity is confined to part of the
homogeneous system. At first sight, this result is very surprising; one might
have expected that the chaotic activity would always spread through the whole system. 

\vspace*{-1cm}
\begin{figure}[htb] 
\begin{picture}(420,0)(0,0)
\end{picture}
\caption{Diffusion coefficient of the non-chaotic and the chaotic pattern for
 $b=0.6$ (other parameters as in fig.\protect{\ref{f:stabi12}}a). 
\protect{\label{f:Dhat}}
}
\end{figure}
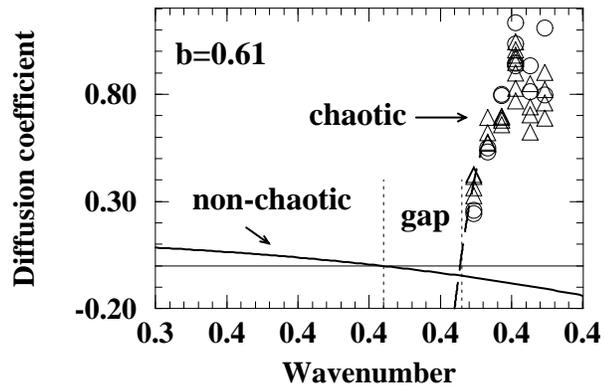

The mechanism for the localization is due to the fact that double phase slips conserve the
total phase, i.e. the wavelength of the pattern is the same before and after the double
phase slip. Thus, a time-averaged pattern has a well-defined wavenumber $\hat{q}$
and a well-defined
phase $\hat{\phi}$. Since the phase is conserved it is expected to 
satisfy a slow evolution equation, which
on symmetry grounds has the form of a diffusion equation
\be
\partial_T \hat{\phi} = \hat{D}(\hat{q}) \partial_X^2 \hat{\phi}.
\ee
Through a detailed numerical study of the response of the extended chaotic state
to a time-periodic perturbation we have explicitly demonstrated the diffusive behavior
of the chaotic state on large scales \cite{GrRi96,GrRi97}. 
This allowed us also to determine the 
effective diffusion coefficient $\hat{D}(\hat{q})$ as a function of $\hat{q}$. The result
is shown in fig.\ref{f:Dhat}. The solid line shows the analytic result for the usual
diffusion coefficient $D(q)$ (cf. (\ref{e:pe})) which becomes negative at the Eckhaus 
stability limit. For larger wavenumbers the pattern is unstable to phase slips and
undergoes persistent double phase slips. However, a state in which this chaotic 
activity is homogeneously distributed over the system is not stable to long-wave perturbations
as long as the effective diffusion coefficient $\hat{D}$ is negative. Thus, there
is a stability gap in wavenumber for which neither the regular nor the chaotic state
are diffusively stable (see fig.\ref{f:Dhat}). 
If the initial wavenumber is chosen in that range the pattern has to
break up into domains with large wavenumber, which are chaotic, and domains
with smaller wavenumber, which are not chaotic. It cannot go into the
chaotic (or the non-chaotic) everywhere in space since the total phase, i.e. the integral
over the wavenumber, is conserved in the process. Thus, if the wavenumber increases in
some part it has to decrease in another part of the system. 
This separation into domains is very much the same as the phase separation found
in equilibrium (fluid) mixtures when they are quenched into the miscibility gap.

\section{Dynamics in Two Dimensions}

The wavelength-changing process that occurs in one dimension $via$ a phase slip
involves in two dimensional patterns the creation (or annihilation) of a pair of
defects (dislocations). It is therefore tempting to speculate that a double phase slip
will correspond to the creation and annihilation of a `bound defect pair', i.e.
two defects will be created together and will annihilate each other soon thereafter. 
This would be in contrast to the dynamics observed usually (e.g. in the single
complex Ginzburg-Landau equation) where the defects that are  created together are
not strongly correlated \cite{GiLe90,RoBo96}. If such a regime of fluctuating bound defect pairs
exists one may expect also a transition in which the defects become unbound. In thermodynamic
equilibrium such unbinding transitions have found great interest in the context of
two-dimensional melting \cite{Ne83} and of vortex unbinding in thin-film superconductors \cite{Mi87}.
We are currently investigating the possibility of such a transition in this non-equilibrium
system.

We have obtained preliminary results that indicate that such a transition may exist 
for parameters corresponding to the stability diagram shown in fig.\ref{f:stabi12}b.
Since no double phase slips seem to occur in the single Ginzburg-Landau equation
describing traveling waves we consider a regime in which there exists a transition
from parametrically forced standing waves to traveling waves, i.e. a parity-breaking
instability. This is the case for $a_r>0$ as shown in fig.\ref{f:stabi12}b. 
For large $b$ the standing waves are unstable at all wavenumbers as in the case
discussed above and double phase slips occur. With decreasing $b$ the parity-breaking
instability is approached and we expect that the double phase slips may become replaced
by single phase slips in its vicinity. 

\begin{figure}[htb] 
\begin{picture}(420,160)(0,0)
\put(-75,-60) {\includegraphics{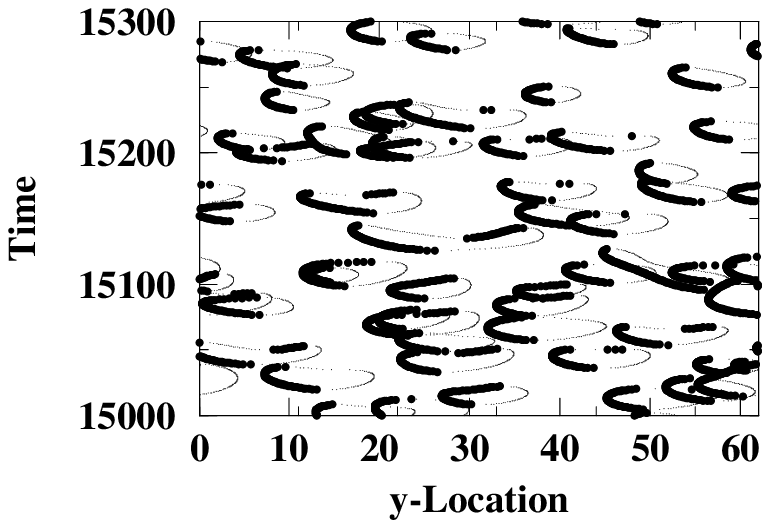}}
\put(165,-60) {\includegraphics{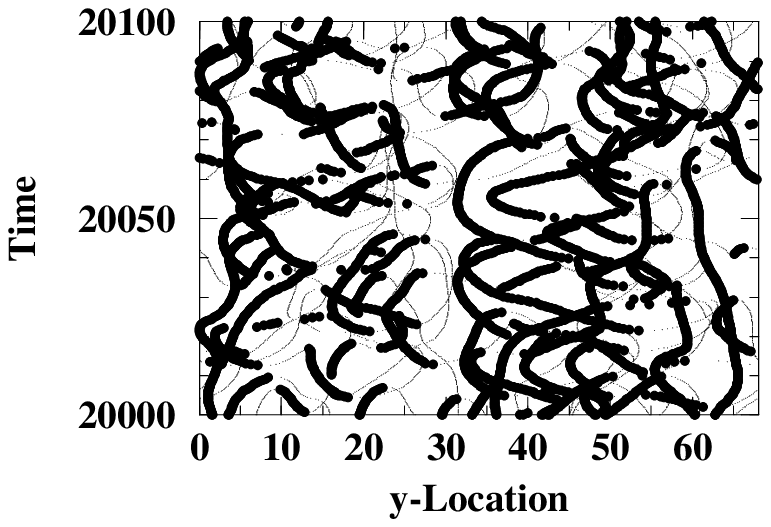}}
\end{picture}
\caption{$y$-component of the trajectory of defects. a) $b=2$ (other parameters as in fig.\protect{\ref{f:stabi12}}a); $b=0.5$  (other parameters as in fig.\protect{\ref{f:stabi12}}b).
\protect{\label{f:loops}}
}
\end{figure}

\begin{figure}[htb] 
\begin{picture}(420,160)(0,0)
\put(-75,-60) {\includegraphics{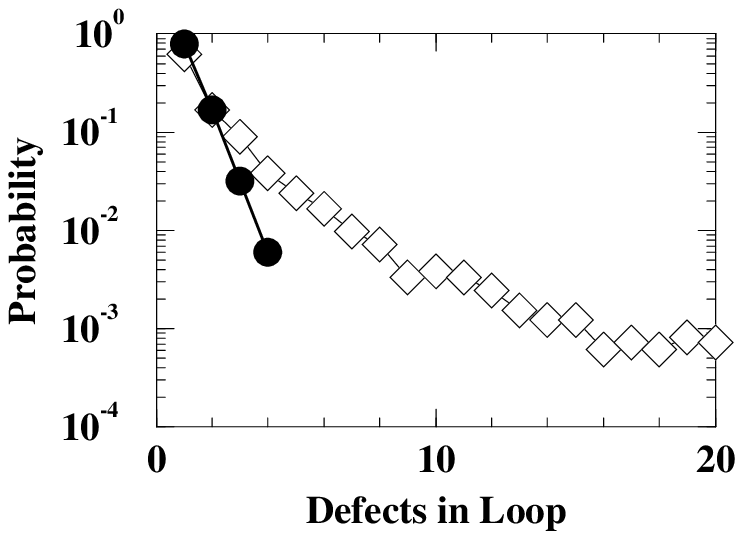}}
\put(165,-60) {\includegraphics{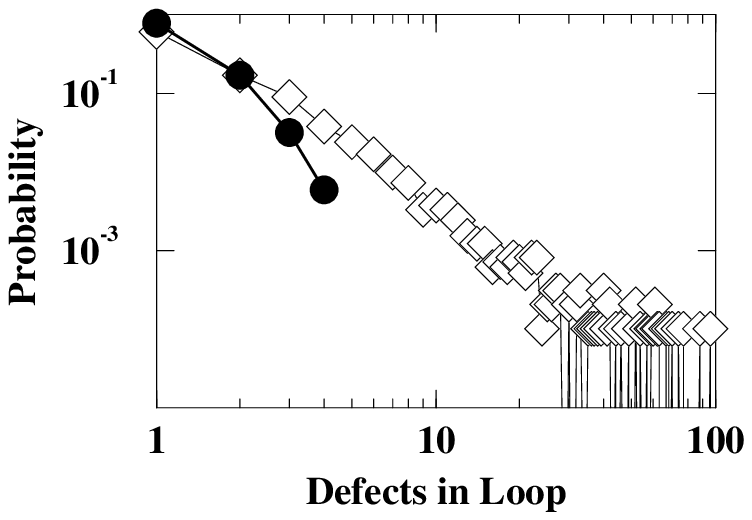}}
\end{picture}
\caption{Relative frequency of defect loops as a function of the number of defect pairs
being part of them. open symbols: $b=2$ (other parameters as in fig.\protect{\ref{f:stabi12}}a); solid symbols: $b=0.5$  (other parameters as in fig.\protect{\ref{f:stabi12}}b). The double-logarithmic plot indicates that for $b=0.5$
the distribution is better approximated by a power law than by an exponential.
\protect{\label{f:loopstat}}
}
\end{figure}

Fig.\ref{f:loops}a,b show space-time diagrams for the $y$-location of defects for simulations
for $b=2$ and $a_r=-0.05$, and $b=0.5$ and $a_r=0.25$. For these simulations (\ref{oneDA},\ref{oneDB})  have  been extended
to two dimensions by replacing the second derivative by a Laplacian. For large $b$ (fig.\ref{f:loops}a)
the defects behave as expected: after their creation they move apart, turn around and
annihilate each other again, forming a loop in the space-time diagram. Even in this regime
there are loops containing more than one defect pair. Fig.\ref{f:loopstat} shows the statistics
of loops of different sizes (solid symbols): 
although larger loops occur, their relative frequency
decays very rapidly (exponentially) with size. For smaller $b$ the space-time diagram
of the defect location is considerably 
more complicated and offers no simple picture of the dynamics. The corresponding
statistics of loop sizes is shown in fig.\ref{f:loopstat} with open symbols. 
Still most loops contain only one defect pair;
this reflects essentially the fact that even if the two defects that are created 
together moved  in a 
completely uncorrelated fashion they would still be annihilated
most likely by their `partner' since it
is closest initially. However, fig.\ref{f:loopstat} shows that large loops are now considerably 
more frequent. In fact, a power-law decay is more consistent with the data than an exponential
decay (cf. fig.\ref{f:loopstat}b). 
 It should be mentioned, that the two-dimensional correlation functions of the patterns
themselves show also a drastic difference: while for $b=2$ the pattern is strongly correlated and
consists of quite ordered stripes, the correlation function decays rapidly and almost isotropically
for $b=0.5$ \cite{GrRi97a}. 
Thus, the two regimes could also be called ordered and disordered defect chaos.

Current and future work is directed to identify whether the change between the two regimes
shown in this paper is smooth or whether it involves a true transition. As diagnostics we 
are not only using the loop-size distribution, but also the defect life-time, the spatial extent of the 
loops, the distances traveled by each defect, as well as correlation functions. 

This work was supported by the United States Department of Energy through 
grant DE-FG02-92ER14303. It also made use of the resources of
the Cornell Theory Center, which receives
major funding from NSF and New York State with additional support from the Advanced Research
Projects Agency, the National Center for Research Resources at the National Institutes
of Health, IBM Corporation and members of the Corporate Research Institute.

\bibliography{/home2/hermann/.bibfiles/journal}

\end{document}